\documentclass[10pt]{raa}            % referee version: for submission
\usepackage[T1]{fontenc}
\usepackage{ae,aecompl}
\usepackage{graphicx,times}             %for PS/EPS graphics inclusion, new
\begin{document}

   \title{Early acceleration of electrons and protons at the nonrelativistic quasiparallel shocks with different obliquity angles
%\footnotetext{$*$ Supported by the National Natural Science Foundation of China.}
}
%   \subtitle{I. Place Your Subtitle Here}

   \volnopage{Vol.0 (20xx) No.0, 000--000}      %%preserved for Editor. DOn't remove!
   \setcounter{page}{1}          %%starting page, preserved for Editor. DOn't remove!

   \author{Jun\ Fang
      \inst{1}
   \and Chun-Yan\ Lu
      \inst{1}
   \and Jing-Wen\ Yan
      \inst{1}
   \and Huan\ Yu
      \inst{2}
   }

   \institute{Department of Astronomy, Key Laboratory of Astroparticle Physics of Yunnan Province, Yunnan University, Kunming 650091, China; {\it fangjun@ynu.edu.cn}\\
        \and
             Department of Physical Science and Technology, Kunming University, Kunming 650214, China; {\it yuhuan.0723@163.com}\\
   }

   \date{xxx}

\abstract{
The early acceleration of protons and electrons in the nonrelativistic collisionless shocks with three obliquities are investigated through 1D particle-in-cell simulations. In the simulations, the charged particles possessing a velocity of $0.2\, c$ flow towards a reflecting boundary, and the shocks with a sonic Mach number of $13.4$ and a Alf\'{v}en Mach number of $16.5$ in the downstream shock frame are generated.
In these quasi-parallel shocks with the obliquity angles $\theta = 15^\circ$, $30^\circ$, and $45^\circ$, some of the protons and the electrons can be injected into the acceleration processes, and their downstream spectra in the momentum space show a power law tail at a time of $1.89\times10^5 \omega_{\rm pe}^{-1}$, where $\omega_{\rm pe}$ is the electron plasma frequency. Moreover, the charged particles reflected at the shock excite magnetic waves upstream of the shock. The shock drift acceleration is more prominent with a larger obliquity angle for the shocks, but the accelerated particles diffuse parallel to the shock propagation direction more easily  to participate in the diffusive shock acceleration. At the time still in the early acceleration stage, more energetic protons and electrons appear in the downstream of the shock for $\theta = 15^\circ$ compared with the other two obliquities; moreover, in the upstream region, the spectrum of the accelerated electrons is the hardest for $\theta_{\rm nB} = 45^\circ$ among the three obliquities, whereas the proton spectra for $\theta_{\rm nB} = 15^\circ$ and $45^\circ$ are similar as a result of the competition of the effectiveness of the shock drift acceleration and the diffusive shock acceleration.
}

   \authorrunning{Fang et al. }            %author_head in even pages
   \titlerunning{Early acceleration of electrons and protons at the nonrelativistic shocks }  % title_head in odd pages

   \maketitle
%% The author head (on even pages) and the title head (on odd pages) will be
%% automatically extracted from \author{} and \title{}. Whenever the title is too long,
%% you will be asked to supply a shorter one by inserting either \authorrunning{} or
%% \titlerunning{} before \maketitle. Anyway, you can specify your own heads.
%%
%%
%% Note: In the following text body of your manuscript, please note several differences from
%%       other major journals:
%% (1) \subsection{Please Capitalize the First Letter of Each Notional Word in Subsection Title}
%% (2) Please Capitalize the First Letter of Each Notional Word in alls' captions

%
%________________________________________________ sections below
%
%-----------------------------------------------------------------------------%
\section{Introduction}
\label{intro}
%-----------------------------------------------------------------------------%

Acceleration of charged particles in nonrelativistic collisionless strong shocks plays an important role in the dynamical evolution of supernova remnant and the origin of the Galactic cosmic rays (Gaisser~\cite{G90}; Orlando et al.~\cite{Oea12}; Fang et al.~\cite{Fea18}; Yu \& Fang~\cite{YF18}).  As illustrated by observations on the supernova remnant evolved in a medium with an uniform magnetic field, such as SN 1006 (Koyama~\cite{K95}; Acero et al.~\cite{Aea10};Vink~\cite{V12}), the distribution of the high-energy particles in the shock varies with the obliquity of the shock.  Particle-in-cell (PIC) and hybrid PIC simulations are usually to be employed to investigate the acceleration process of the high-energy particles (Yang et al.~\cite{Yea09}; Guo \& Giacalone~\cite{GG15}; Kato~\cite{K15}; Park et al.~\cite{Pea15}; Matsumoto et al.~\cite{Mea17}). In the quasi-parallel shocks with the obliquity angle, i.e., the angle between the shock normal and the undisturbed background magnetic field, $\theta_{\rm nB}=30^{\circ}$,  using 1D PIC simulations, the former studies in Park et al.~(\cite{Pea15}) and Kato~(\cite{K15}) indicated that both protons and electrons could be injected to the diffusive acceleration process after preheating, and the particle spectra had power-law distributions in the high-energy tail. Especially, the index of $\sim -4$ for the power law distribution of the accelerated particles in the momentum space was indicated from the simulations given in Park et al.~(\cite{Pea15}), which is consistent with the prediction of the diffusive shock acceleration for nonrelativistic strong shocks. Alternatively, for the quasi-perpendicular shock with an obliquity angle of $\theta_{\rm nB}=45^{\circ}$, using 2D3V PIC simulations, Wieland et al.~(\cite{Wea16}) obtained two distinctive shocks and a contact discontinuity by initiating the plasma with two components with different densities and velocities as in Niemiec~(\cite{Nea12}). Their results indicated that the ion spectra had a supra-thermal tail, whereas there was no indication of acceleration of the electrons in the quasi-perpendicular shocks.

Electron acceleration involved in low Mach number shocks has also been investigated via PIC simulations. For a quasi-perpendicular shock with a low Mach number, electrons can interact with the magnetic wave generated upstream, and they can be pre-accelerated due to repeated cycles of the shock drift acceleration (SDA) (Guo et al.~\cite{Gea14a,Gea14b}).
Guo et al.~(\cite{Gea14b}) explored electron acceleration at low Mach number shocks, which usually appeared in galaxy clusters and solar flares, using 2D PIC simulations.  Their results indicated that the electrons upstream of the shock could be scattered with the oblique magnetic waves generated by the escape of the electrons ahead of the shock via the firehose instability. Electrons can be accelerated via shock drift acceleration when bouncing between the shock and the upstream region. Ha et al.~(\cite{Hea18}) studied the injection of protons into the diffusive shock acceleration (DSA) for the quasi-parallel shocks with low sonic Mach numbers ($M_{\mathrm{s}}$) and a high plasma beta of  $\sim 100$, and they found that only those shocks with $M_{\mathrm{s}}\geq 2.25$ could result in an efficient injection. Moreover, Yang et al.(\cite{Yea18}) investigated the electron reflection at parallel and perpendicular shocks with low Mach numbers using 2D PIC simulatons, and the results showed that the reflected electron beam can be produced as a result of the micro-turbulence at the shock foot. 

Besides PIC simulations, hybrid PIC simulations, which treat ions kinetically with electrons modeled as a charge-neutralizing fluid to reduce the computational expense, are also performed to study the acceleration process of the ions in shocks (Gargat\'{e} \& Spitkovsky~\cite{GS12}; Caprioli \& Spitkovsky~\cite{CS14a,CS14b}; Hao et al.~\cite{Hea16}). For the strong parallel/quasi-parallel shocks, the proton acceleration is efficient, and the power law spectra have a dependence of $dN/dp \propto p^{-4}$ (Caprioli \& Spitkovsky~\cite{CS14a}). However, for the quasi-perpendicular ones, the acceleration is insufficient because the protons only gain a factor of a few in momentum before they convect away from the shock (Caprioli \& Spitkovsky~\cite{CS14a}).  Derived from the 2D hybrid PIC simulations given by Hao et al.(~\cite{Hea16}), ions flow across the shock from upstream to downstream more easily in the upper part of the shock, whereas the upstream ions tend to be reflected in the lower part.

In this paper, we intend to investigate the acceleration effectiveness of the protons and the electrons in the nonrelativistic collisionless shocks with different obliquity angles using 1D3V PIC simulations, and then the dependence of effectiveness on the obliquity can be derived.
Our work is different with the previous works in Park et al.~(\cite{Pea15}) and Kato~(\cite{K15}) because they investigated the acceleration processes in the shock with one obliquity with $\theta_{\rm nB}=30^{\circ}$ with Mach numbers of several tens. Although the shocks with other obliquity angles have been studied using PIC simulations, the dependence of the acceleration effectiveness of both protons and electrons on the obliquity of the shock cannot be integrated from these papers due to the difference of the shock parameters or the dimension used in the different papers. For example, the weak shocks with the much smaller sonic Mach number $\leq 4$ are investigated in Ha et al.~(\cite{Hea18}). Moreover, in Wieland et al.~(\cite{Wea16}), the simulation is performed in 2D3V up to a time much smaller than that in Park et al.~(\cite{Pea15}).  The simulation setup is presented in Section \ref{simuset}. The results from the simulations are described in Section \ref{resu}. Finally, we show a summary and discussion in Section \ref{sumdis}, .
%-----------------------------------------------------------------------------%
\section{Simulation setup}
\label{simuset}
Simulations for the collisionless shocks are performed in 1D3V, i.e., 1D along $x$ with all components of velocity and field retained, using the parallel electromagnetic PIC code (Derouillat et al.~\cite{Dea18}). Initially, the plasma consisted of electrons and protons with a bulk velocity of $\textbf{v}=-v_0 \hat{\textbf{x}}$ against a reflecting wall at $x=0$.  Our simulations are performed in the the downstream frame, and $v_0 = 0.2 c$, where $c$ is the speed of light, and a reduced proton-to-electron mass ratio $m_{\rm p}/m_{\rm e} = 30$ is adopt to save the computational resources.

In the simulations, the initial magnetic field is $\textbf{B}_{\rm i}=B_0 (\cos \theta_{\rm nB} \hat{\textbf{x}} + \sin \theta_{\rm nB} \hat{\textbf{y}})$,  and the temperatures of the electrons and the protons are $T_{\rm e} = T_{\rm p} = 0.2\times 10^{-3} m_e c^2/k_{\mathrm{B}}$, where $\theta_{\rm nB}$ is the angle between the shock normal and the initial magnetic field, $m_e$ is the electron mass, and $k_{\mathrm{B}}$ is the Boltzmann constant. The sound speed of the upstream bulk flow is $c_{\rm s} = \sqrt{2 \gamma_{\rm ad}k_{\rm B}T_{\rm p}/ m_{\rm p} }$, where $\gamma_{\rm ad}= 5/3$ is the adiabatic index for the nonrelativistic shock. For the compression ratio $r=4$ and $n_0=0.1\,{\rm cm}^{-1}$, the sonic and Alf\'{v}en Mach numbers in the downstream frame are $M_{\rm s} =  v_{\rm sh}/c_{\rm s} = 13.4$ and $M_{\rm A} =  v_{\rm sh}/v_{\rm A}=16.5$, respectively, where $v_{\rm sh}=v_0 r/(r-1)$, $v_{\rm A} = B_0/\sqrt{4\pi n_0 m_{\rm p}}$ is the Alf\'{v}en speed, and $n_0$ is the number density of the protons in the initial plasma. The simulation results are indicated in units of $c/\omega_{\rm pe}$ and $\omega_{\rm pe}^{-1}$ for length and time, respectively, where  $\omega_{\rm pe}=\sqrt{4\pi e^2 n_0/m_e}$ is the electron plasma frequency. The variation scale of the shock structure is determined by the proton Larmor radius $r_{{\rm L, p}}= m_{\rm p} v_0 c/(eB_0)= 67.6 c/\omega_{\rm pe}$ and the proton gyration period $\Omega_{\rm cp}=m_{\rm p} c/(eB_0)=338 \omega_{\rm pe}^{-1}$. The grid resolution is $\Delta x=0.1 c/\omega_{\rm pe}$, and the time step is $\Delta t=0.045 \omega_{\rm pe}^{-1}$ with 64 particles per cell per species and the $x$ dimension stretching to $6.4\times 10^4 c/\omega_{\rm pe}$.

\section{Results}
\label{resu}
\subsection{Shock structures and particle energy spectra}
\label{shockstructure}

\begin{figure}[tbh]
\begin{center}
\includegraphics[width=0.5\textwidth]{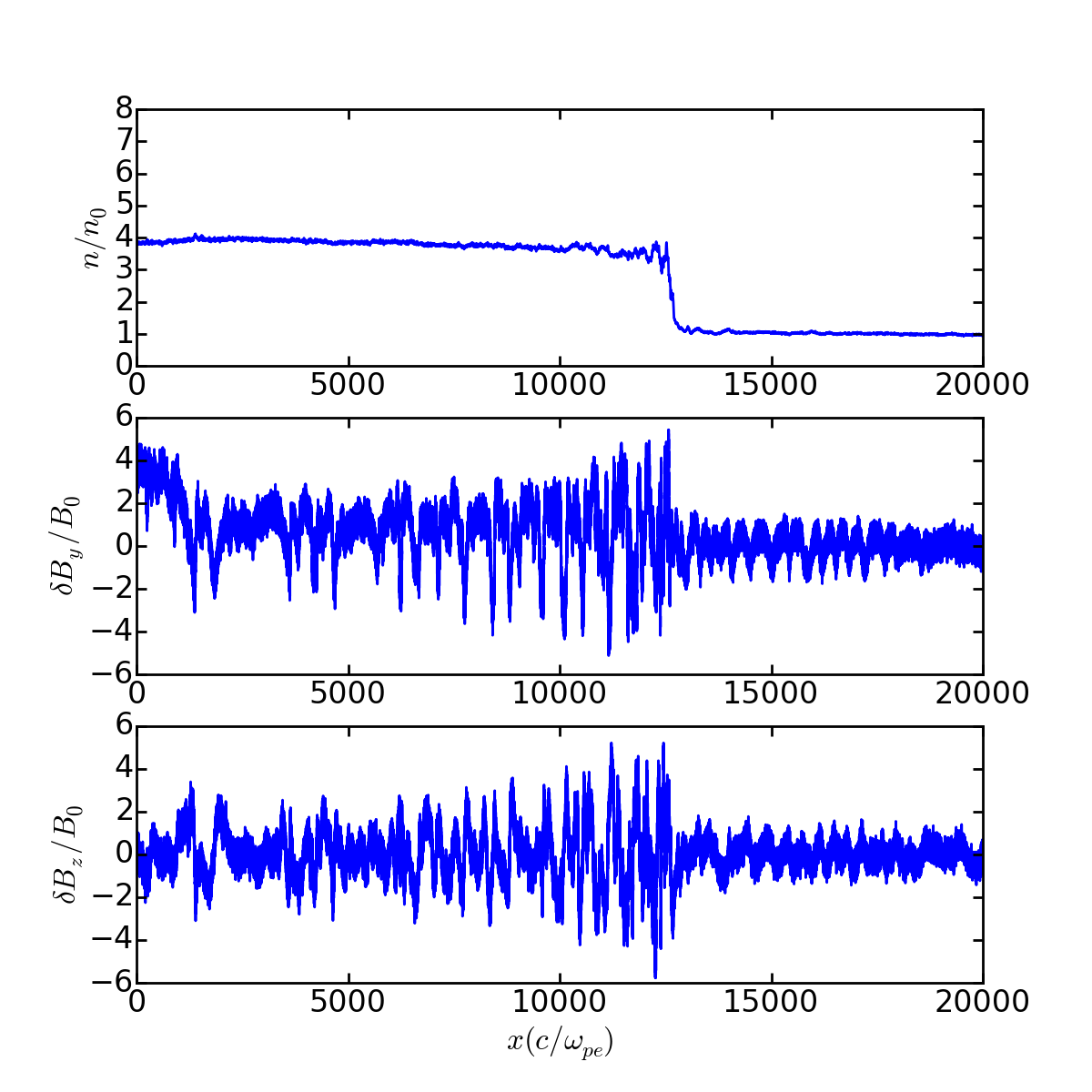}
\end{center}
\caption{Proton density profile (upper panel) and transverse magnetic fields $\delta B_y$ (middle panel) and $\delta B_z$ (lower panel) at $t  = 1.89\times10^5 \omega_{\rm pe}^{-1}$ for the shock with  $T_{\rm e} = T_{\rm p} = 0.2\times 10^{-3} m_e c^2/k_{\mathrm{B}}$, $v_0 = 0.2 c$, $m_{\rm p}/m_{\rm e} = 30$ and $\theta_{\rm nB}=15^{\circ}$.}
\label{fig:nb}
\end{figure}

\begin{figure}[tbh]
\begin{center}
\includegraphics[width=0.5\textwidth]{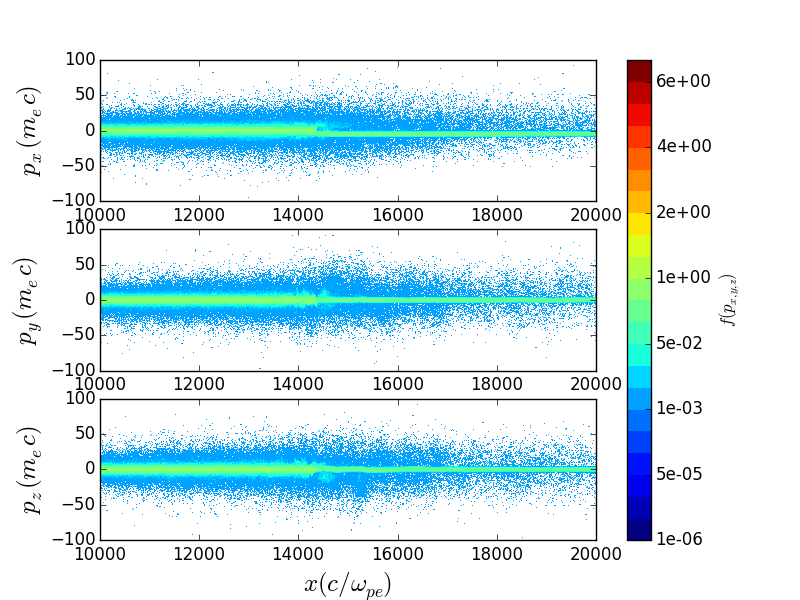}
\end{center}
\caption{The momentum phase space plots for the protons in the shock with the same parameters  as Figure\ref{fig:nb}. }
\label{fig:protondis}
\end{figure}

Figure\ref{fig:nb} shows the profile of the proton number density normalized by the far upstream value for a quasi-parallel shock with $\theta_{\rm nB}=15^{\circ}$ at $t=1.89\times10^5 \omega_{\rm pe}^{-1}$. Although oscillation occurs around the shock at $\sim 1.25\times10^4 c/\omega_{\rm pe}$, the density is compressed by the shock with a mean compression ratio of $r=4$ for the far downstream particles, which is consistent with that derived by the Rankine-Hugoniot relation, i.e.,
\begin{equation}
  \label{eq:rratio}
  r = \frac{\gamma_{\rm ad}+1}{\gamma_{\rm ad} - 1 + 2/M_{\rm s}}.
\end{equation}
Around the shock, the electrons diffuse more easily from downstream  to upstream than the protons, and the downstream medium, where net positive charged particles accumulate, has a positive potential relative to the upstream one (Gedalin \& Balikhin~\cite{GB04}; Amano, \& Hoshino~\cite{AH07}; Marcowith et al.~\cite{Mea16}). As illustrated in the top panel of Figure \ref{fig:protondis}, a fraction of the protons have been reflected by the shock to possess a positive momentum along $+x$ direction in the upstream region. The other protons impinging on the shock  advect downstream after thermalized in the shock transition zone.

\begin{figure}[tbh]
\begin{center}
\includegraphics[width=0.5\textwidth]{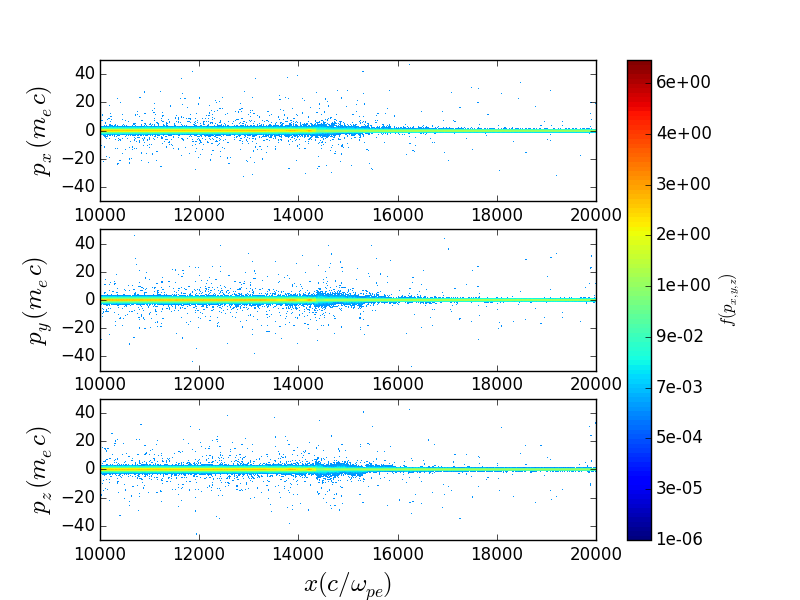}
\end{center}
\caption{The momentum phase space plots for the electrons in the shock with the same parameters as Figure\ref{fig:nb}.}
\label{fig:electrondis}
\end{figure}

As the reflected protons propagating upstream, waves are generated  to greatly amplify the transverse magnetic field, which is also compressed by the shock due to the freezing of magnetic flux in the plasma (Guo et al.~\cite{Gea14a}). As the local magnetic field changing from quasi-parallel to quasi-perpendicular, protons can also be reflected due to the effect of magnetic mirror in the foreshock region (Sundberg et al.~\cite{Sea16}). In SDA, a proton which encounters the shock is accelerated by the motional electric field. This process can last several gyrocycles until the proton gains enough energy to diffuse across the shock and be injected into the DSA, through which the particle gains energy due to collision with the upstream waves (Park et al.~\cite{Pea15}). Figure \ref{fig:protondis} indicates that some protons with high momentum diffused into the upstream region far away from the shock with a distance much larger than their gyroradiuses.

Different from the protons, the electrons that encounter the shock can be reflected to the upstream region only through magnetic mirror, because the shock potential intends to pull them downstream of the shock. The reflected electrons are scattered by the upstream waves, and they  gyrate along the upstream magnetic field. As illustrated in Figure\ref{fig:electrondis}, they have large and positive $p_x$ and $p_y$. These reflected electrons can be effectively accelerated via SDA when they are trapped closely to the shock due to magnetic mirror at the shock and scattering with the upstream waves. Moreover, the electrons which gain enough energy can also be injected into the DSA process to be further accelerated (Park et al.~\cite{Pea15}).

\begin{figure}[tbh]
\begin{center}
\includegraphics[width=0.5\textwidth]{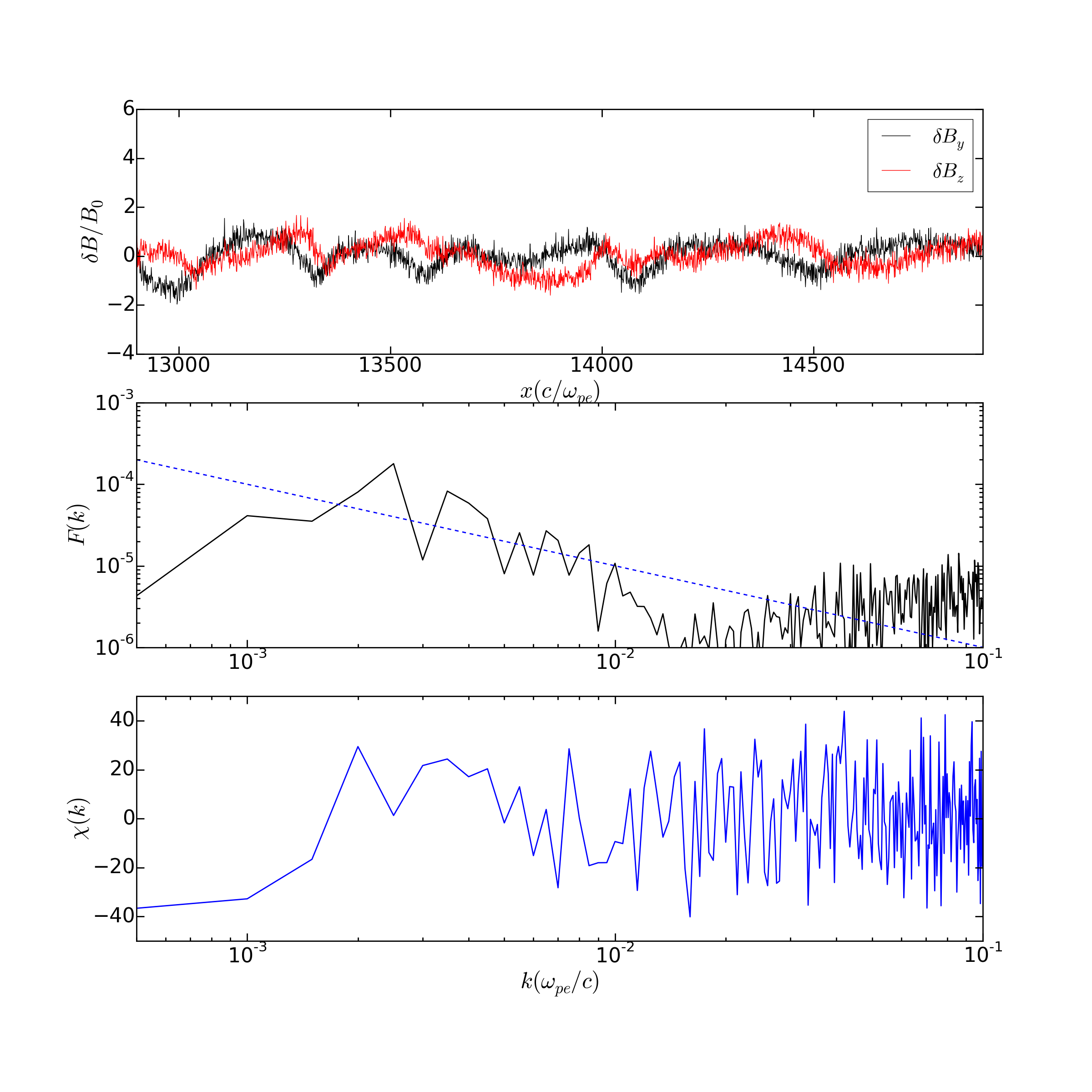}
\end{center}
\caption{Spacial and spectral distributions of the self-excited transverse magnetic field. From top to bottom, the ratios of $\delta B_y / B_0$ and $\delta B_z / B_0$ on $x$, spectral distributions of $F(k)$, and the Polarization angle $\chi(k)$ on $k$ are presented. In the middle, the dashed line represents the power law distribution, $F(k)\propto k^{-1}$.}
\label{fig:byfft}
\end{figure}

As the reflected energetic protons stream upstream along the magnetic field for the quasi-parallel shock with $\theta_{\rm nB}=15^{\circ}$, magnetic turbulence is excited due to CR-driven instabilities. Moreover, left-handed polarized waves are excited via resonant instabilities, whereas right-handed waves are produced through nonresonant instabilities (Caprioli \& Spitkovsky~\cite{CS14b}). Figure \ref{fig:byfft} indicates the spatial distribution of the excited magnetic field $\delta B_y = B_y - B_0 \sin \theta_{\rm nB}$ and $\delta B_z$  in units of $B_0$ and Fourier analyses of these magnetic fields in the precursor region with $x$ between $1.29\times10^4 c/\omega_{\rm pe}$ and $ 1.49\times10^4 c/\omega_{\rm pe}$.  The distribution of the magnetic energy density for the self-generated magnetic field per unit logarithmic bandwidth of waves in $k$ space is indicated by $F(k) = k(|\delta B_y(k)|^2 + |\delta B_z(k)|^2)/B_0^2$ (Caprioli \& Spitkovsky~\cite{CS14b}). The process of the resonant streaming instability with nonthermal particles, which has a distribution of $f(p)\propto p^{-4}$ for a strong shock, can excite the upstream Alfv\'{e}n waves with $F(k) \propto k^{-1}$ \cite{CS14b}. The middle panel in Figure \ref{fig:byfft} shows that $F(k)$ dominates in the region $2\times 10^{-3} \omega_{\rm pe}/c \leq k \leq 2\times 10^{-2} \omega_{\rm pe}/c$, and the spectrum is relatively steeper than $F(k) \propto k^{-1}$. Furthermore, the spectrum with $k>0.02 \omega_{\rm pe}/c$ is harder than the power-law distribution.

The handedness of waves can be inferred from $\chi$ with positive (negative) value representing right-handed (left-handed) polarization (Park et al.~\cite{Pea15}; Ha et al.~\cite{Hea18}). Figure \ref{fig:byfft} also shows the polarization angle,
\begin{equation}
  \label{eq:chi}
  \chi(k) = \frac{1}{2}\sin^{-1}\left(\frac{V}{I}\right),
\end{equation}
where $I(k)=|\delta B_y(k)|^2 + |\delta B_z(k)|^2$ and $I(k)=|\delta B_y(k)|^2 - |\delta B_z(k)|^2$  (Ha et al.~\cite{Hea18}).
With $k>2\times 10^{-3} \omega_{\rm pe}/c$, although the waves with both positive and negative $k$ are excited, the modes with positive $k$ in more frequent. Taking the distribution of $F(k)$ and $\chi(k)$ into account, the nonresonant instabilities are more effective at $t=1.89\times10^5 \omega_{\rm pe}^{-1}$ in the simulation.

\begin{figure}[tbh]
\begin{center}
\includegraphics[width=0.5\textwidth]{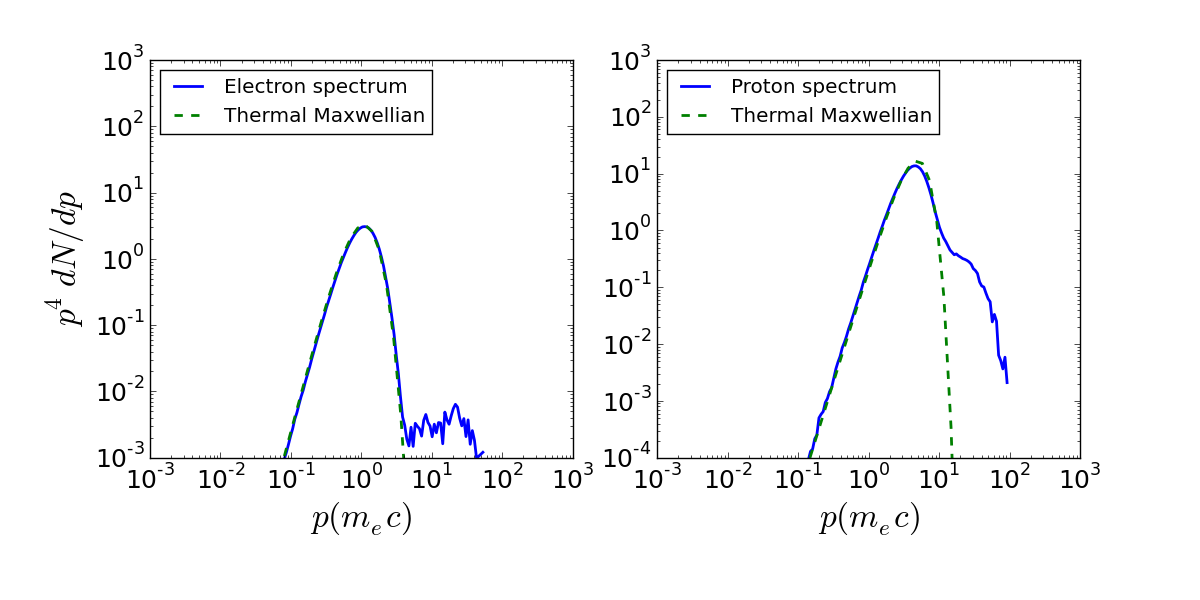}
\end{center}
\caption{Downstream spectra of electrons (left panel) and protons (right panel). The thermal Maxwellian distributions with $kT_e= kT_p= 0.2m_e c^2$ are also indicated as the dashed lines for the electrons and the protons, respectively.}
\label{fig:accpar}
\end{figure}

Figure \ref{fig:accpar} shows the downstream spectra in the momentum space for protons and electrons with $x$ ranging from $1.0\times10^4 c/\omega_{\rm pe}$ to $1.2\times10^4 c/\omega_{\rm pe}$, and the dashed lines represent the thermal Maxwellian distribution with $kT=0.2m_e c^2$. A nonthermal tail, which represents the accelerated particles via SDA and DSA, is attached on the thermal distribution both for the electrons and for the protons. For the electrons, the turning momentum from the Maxwellian distribution to the power law is $\sim 5 m_e c$, and the cut off momentum, which is limited by the simulation time, is about $50 m_e c$. For the protons, the high-energy tail extends from $\sim 15 m_e c$ to $\sim 100 m_e c$.

\subsection{Dependence on the obliquity angle}
%-----------------------------------------------------------------------------%

\begin{figure}[tbh]
\begin{center}
\includegraphics[width=0.5\textwidth]{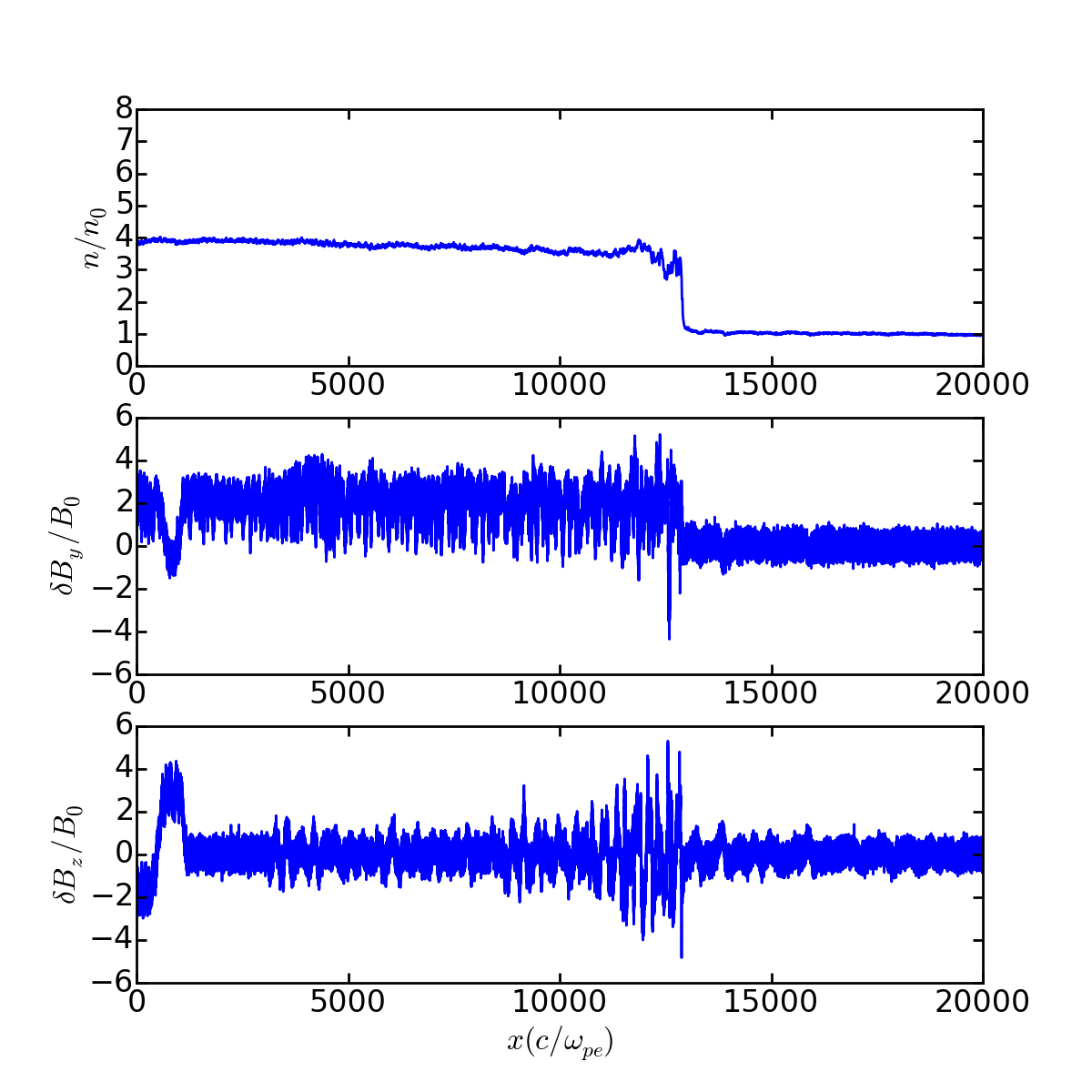}
\end{center}
\caption{Density profile of the protons (upper panel) and transverse magnetic fields $\delta B_y$ (middle panel) and $\delta B_z$ (lower panel) at $t  = 1.89\times10^5 \omega_{\rm pe}^{-1}$ for the shock with  $T_{\rm e} = T_{\rm p} = 0.2\times 10^{-3} m_e c^2/k_{\mathrm{B}}$, $v_0 = 0.2 c$, $m_{\rm p}/m_{\rm e} = 30$ and $\theta_{\rm nB}=45^{\circ}$.}
\label{fig:nb60}
\end{figure}

The shock structure and the involved acceleration processes of protons and electrons at the shock depend on the obliquity of the shock.
Figure \ref{fig:nb60} shows the proton density profile and the transverse magnetic field for the  shock with $\theta_{\rm nB}=45^{\circ}$. Similar as $\theta_{\rm nB}=15^{\circ}$, the density oscillates at the shock, and in the far downstream it is $4$ times of the upstream one. The magnetic field upstream of the shock is exited with smaller wavebands compared with $\theta_{\rm nB}=15^{\circ}$.

\begin{figure}[tbh]
\begin{center}
\includegraphics[width=0.5\textwidth]{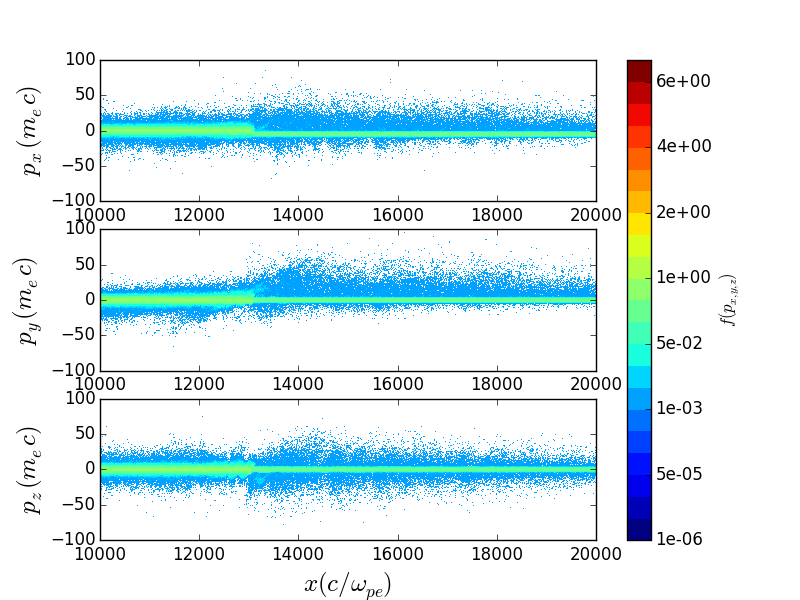}
\end{center}
\caption{The momentum phase space plots for the protons in the quasi-perpendicular shock with $\theta_{\rm nB}=45^{\circ}$. }
\label{fig:protondis_60}
\end{figure}

\begin{figure}[tbh]
\begin{center}
\includegraphics[width=0.5\textwidth]{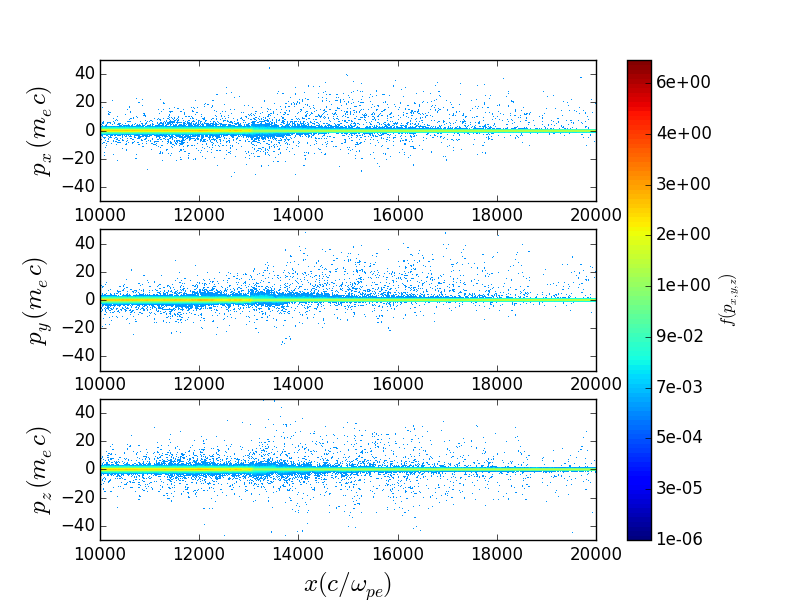}
\end{center}
\caption{The momentum phase space plots for the electrons in the quasi-perpendicular shock with $\theta_{\rm nB}=45^{\circ}$.}
\label{fig:electrondis_60}
\end{figure}

Figure \ref{fig:protondis_60} and Figure \ref{fig:electrondis_60} show the phase space plots $p_x-x$, $p_y-x$, $p_z-x$ for the protons and the electrons in the shock with $\theta_{\rm nB}=45^{\circ}$, respectively. The protons and electrons gyratint along the magnetic field perpendicular to the shock can accumulate energy via the shock drift acceleration. As illustrated in the figures, most of the reflected protons and electrons gyrate along the upstream magnetic field with positive $p_x$ and $p_y$. Compared with $\theta_{\rm nB}=15^{\circ}$, the accelerated particles are reflected by the shock more effectively for $\theta_{\rm nB}=45^{\circ}$, and more high-energy particles are confined in the upstream region.

\begin{figure}[tbh]
\begin{center}
\includegraphics[width=0.5\textwidth]{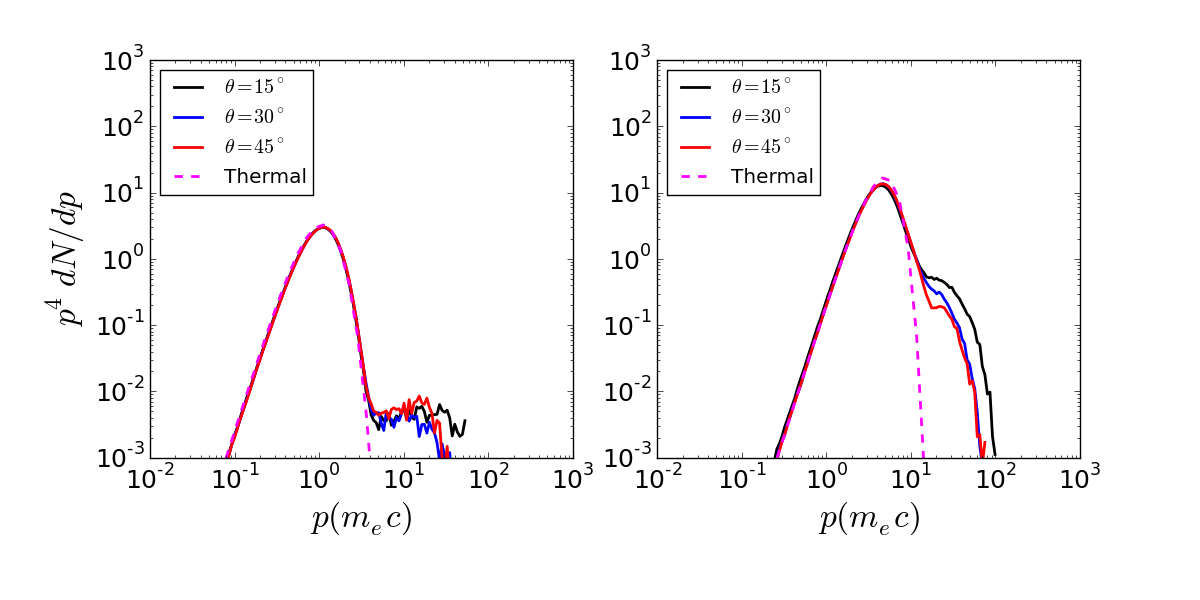}
\end{center}
\caption{Downstream spectra of the  electrons (left panel) and the  protons (right panel) with $1\times10^4 c/\omega_{\rm pe} \leq x \leq 1.2\times10^4 c/\omega_{\rm pe}$ for the shock with $\theta_{\rm nB}=15^{\circ}$, $30^{\circ}$, and $45^{\circ}$. The thermal Maxwellian distributions with $kT_e= kT_p= 0.2m_e c^2$ are also indicated as the dashed lines for the electrons and the protons, respectively.}
\label{fig:parspectra_down}
\end{figure}

\begin{figure}[tbh]
\begin{center}
\includegraphics[width=0.5\textwidth]{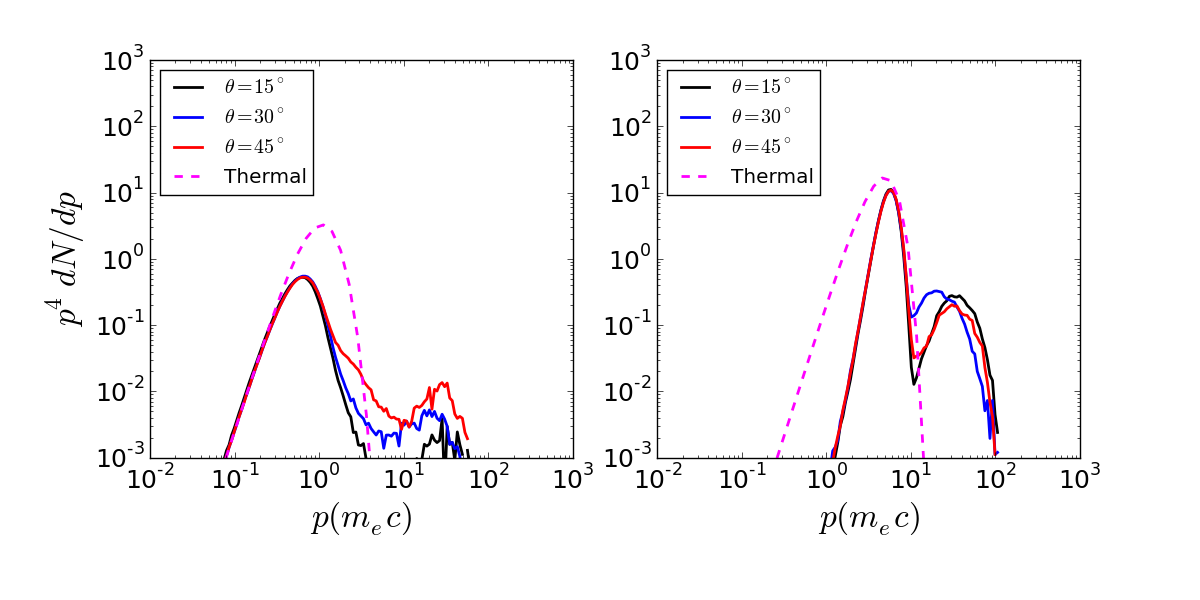}
\end{center}
\caption{Upstream spectra of the electrons (left panel) and the  protons (right panel) with $1.3\times10^4 c/\omega_{\rm pe} \leq x \leq 1.5\times10^4 c/\omega_{\rm pe}$ for the shock with $\theta_{\rm nB}=15^{\circ}$, $30^{\circ}$, and $45^{\circ}$. The others are the same as Figure \ref{fig:parspectra_down}.}
\label{fig:parspectra_up}
\end{figure}

Figure \ref{fig:parspectra_down} shows the particle spectra in the downstream region with $1\times10^4 c/\omega_{\rm pe} \leq x \leq 1.2\times10^4 c/\omega_{\rm pe}$ for the shock with the three obliquity angles, and the spectra in the upstream region $1.3\times10^4 c/\omega_{\rm pe} \leq x \leq 1.5\times10^4 c/\omega_{\rm pe}$ are indicated in Figure \ref{fig:parspectra_up}. As illustrated in the left panel of Figure \ref{fig:parspectra_up}, the SDA is more efficient with a larger obliquity angle for the shocks. Alternatively, in the shock with a smaller obliquity angle, the accelerated particles diffusive more easily in the propagation direction of the shock to take part in the DSA. 
Compared with the protons, the accelerated electrons have lower momentum, and they are more easily confined near the shock to take part in the shock drift acceleration.  Therefore, the protons and the electrons behave differently depending on the obliquities of the shocks. Therefore, the upstream spectra of the protons for $\theta_{\rm nB}=15^{\circ}$  are similar as $45^{\circ}$ due to the competition between the SDA and the DSA (see the right panel of Figure \ref{fig:parspectra_up}). Moreover, the accelerated particles flow from upstream to downstream more easily in the shock with a small obliquity, and the nonthermal spectra for both the protons and the electrons downstream of the shock extend to higher energies for $\theta_{\rm nB}=15^{\circ}$ compared with the other two obliquities (see Figure \ref{fig:parspectra_down}). Especially, the indexes of the nonthermal tails in the electron spectra downstream of the shocks are about $4$ in the momentum space.

\section{Summary and discussion}
\label{sumdis}
In this paper, the early acceleration of the protons and the electrons in nonrelativistic shocks with three obliquity angles are investigated using  1D3V PIC simulations. Some of the charged particles are reflected from the shock, and the magnetic field upstream of the shock is significantly amplified due to CR-driven instabilities. For the three obliquities,
a part of the charged particles are accelerated by the shock, and the spectra of both the protons and the electrons show nonthermal tails.

As illustrated in Park et al.~\cite{Pea15}, some of the charged particles encountered by the shock can be accelerated by the SDA, and they can diffuse across the shock to be injected in the DSA if enough energies are accumulated. Among the three obliquities, the shock drift acceleration is the most effective for $\theta_{\rm nB}=45^{\circ}$. However, the particles diffuse across the shock more easily for a smaller obliquity. As a result, more energetic protons and electrons are distributed in the downstream of the shock with $\theta_{\rm nB}=15^{\circ}$.  There are more energetic electrons in the upstream of the shock with $\theta_{\rm nB}=45^{\circ}$ due to the effectiveness of the shock drift acceleration, and the proton spectra upstream of the shock for $\theta_{\rm nB}=15^{\circ}$ are similar as $45^{\circ}$ resulting from the competition of the effectiveness of the shock drift acceleration and the diffusive shock acceleration.

The simulations in this paper are just performed up to a time of $t  = 1.89\times10^5 \omega_{\rm pe}^{-1}$. At this time, most of the accelerated particles are involved in the SDA, and the DSA is not well developed. We will use longer simulations with more extension region to investigate the dependence of the diffusive shock acceleration in nonrelativistic shocks on the obliquity in the future.

\begin{acknowledgements}
JF is partially supported by the National Key R\&D Program of China under grant No.2018YFA0404204, the
Natural Science Foundation of China (NSFC) through grants 11873042 and 11563009,
the Yunnan Applied Basic Research
Projects under (2016FB001, 2018FY001(-003)), the Candidate Talents Training Fund of Yunnan Province (2017HB003)
and the Program for Excellent Young Talents, Yunnan University (WX069051, 2017YDYQ01). HY is partially supported by the Yunnan Applied
Basic Research Projects (2016FD105), the foundations of Yunnan Province (2016ZZX180, 2016DG006) and Kunming University (YJL15004, XJL15015).
\end{acknowledgements}

\label{lastpage}

\end{document}